\documentclass[doublecol]{epl2} 

\title{CdS/CdSe/CdS spherical quantum wells as single-photon sources}
\shorttitle{CdS/CdSe/CdS spherical quantum wells as single-photon sources} 

\author{A.~Allemand\inst{1}, F.~Kulzer\inst{1}, B.~Mahler\inst{1}, C.~Dujardin\inst{1} and J.~Houel\inst{1,*}}
\shortauthor{A.~Allemand \etal}

\institute{                    
  \inst{1} Univ. Lyon, Université Claude Bernard Lyon 1, CNRS, Institut Lumière Matière, 69622 Villeurbanne, France
}

\abstract{We have synthesized CdS(1.3\,nm)/CdSe(1.7\,nm)/CdS(3.4\,nm) spherical quantum wells with 13\,nm diameter and demonstrated the first antibunching of their emission, labelling them as single-photon sources. Antibunching survives even at high excitation intensities, ruling-out any bi-exciton emission. For the largest intensities, antibunching coupled to spectral measurements, reveal the signature of a blue-shifted emission, associated to an irreversible photo-aging effect. We demonstrate a moderate correlation between the main and the blue-shifted emission energy. Intensity-timetraces show low-blinking, with a median time spent in the bright state of 89\,$\%$. Emission lifetime measurements reveal a complex emission dynamic with either three or four components.  While spherical quantum wells have been initially designed for laser-oriented applications, we demonstrate that they can serve as single-photon sources.}
\begin{document}

\maketitle

\section{Introduction}

Colloidal quantum dot based on CdSe~\cite{murray93sep22} have been at the state-of-the-art of colloidal nano-objects for numerous applications, such as lasers~\cite{klimov_science_2000} and LEDs~\cite{colvin94LED,shirasaki13NatPhot_LEDs,shen19NatPhot_LEDs}, biological trackers~\cite{alivisatos04jan,alivisatos05annur755}, and more recently spectrometers~\cite{bao15nature_qdpsectro}. They were praised for their remarkable optical properties (fast recombination lifetime~\cite{lounis00oct27,lounis03PRL}, single photon emission~\cite{lounis00oct27, vukovic12SPS}, high emission quantum yield at room temperature~\cite{ebenstein02may27,mcbride06QY,klimov11PRL}, narrow emission linewidth~\cite{nirmal96oct31,chen13NatMat_linewidth,fernee13nanotech}) in the early 2000s for applications in nano-photonics~\cite{belacel13nanolett_patch}. However, their drawbacks, namely fluoresence intermittency, photobleaching and fragility against encapsulation in inorganic compound have hindered their potential~\cite{vansark01bleaching,efros16natna11661}. Over the past decade, the road towards non-blinking, robust and high emission quantum-yield colloidal quantum dots have been filled with success~\cite{mahler08natma7659}. While the work-horse of CdSe-based colloidal nanocrystals is the core/shell structure, with a core made of CdSe and a shell constituted of higher bandgap materials such as ZnS, CdS and ZnSe, there have been other marginal, but nonetheless successful, interesting structures~\cite{battaglia03angch_quantumshell,pang05tetrapod,salini14multishell}. One of them is the spherical quantum well (SQW) consisting of first a CdS core and a CdSe shell serving as active medium~\cite{du10nanoshell}, and a covering shell leading to CdS/CdSe/CdS structures~\cite{battaglia03angch_quantumshell,xu05nanoshell,jeong16acsna109297}. From the band structure, it has been assumed and theoretically argued that the confinement is such as for a quantum-well~\cite{zamkov17JACS_1D_SQW}, leaving-out potential single photon applications. CdS/CdSe/CdS SQWs have been recently developed to solve the problem of emission quantum yield decreasing with increasing emission wavelength or shell thickness on CdSe/CdS core/shell standard structures~\cite{pal12nanoletters_LEDs_shell_thickness}. The resulting CdS/CdSe/CdS SQWs were exhibiting reduced blinking (spending $\sim$90\,$\%$ of their emission time in the bright state), robust and fast emission with close to 100\,$\%$ emission quantum yield at room temperature with at diameter up to 15\,nm~\cite{jeong16acsna109297}. Since then, CdS/CdSe/CdS SQWs have demonstrated their use in different applications~\cite{song18nanolett_solar_concentraor_cdse,zamkov20nanoshell_bohr,meng21nanoscale}. However, while CdSe/CdSe/CdS SQWs have seen their optical properties probed extensively at the ensemble levels, there is a lack of statistically significant experiments at the individual SQW level.
In this article, we report on the optical characterization of individual CdS/CdSe/CdS SQWs emitting at 650\,nm. We first demonstrate that the SQWs emit single photons and that the confinement of the excitons is 3-dimensional, contrary to what has been argued in the literature~\cite{zamkov17JACS_1D_SQW,zamkov20nanoshell_bohr}. We then present spectral measurements on single SQWs which demonstrate blue-shifted, size-correlated, irreversible photo-aging. Furthermore, we present single SQWs intensity timetraces which exhibit a low two-level blinking with at least one bright state and a dimmed state. Finally, lifetime measurements exhibit a more complex dynamics of the SQW decays than is usually reported~\cite{jeong16acsna109297}, with multi-exponential behavior demonstrating that note only one, but three linear excitons entities are emissive in SQWs, and no bi-exciton are observed. 

\section{Results and Discussion}
\begin{figure}[t]
\centering
\includegraphics[width=8 cm]{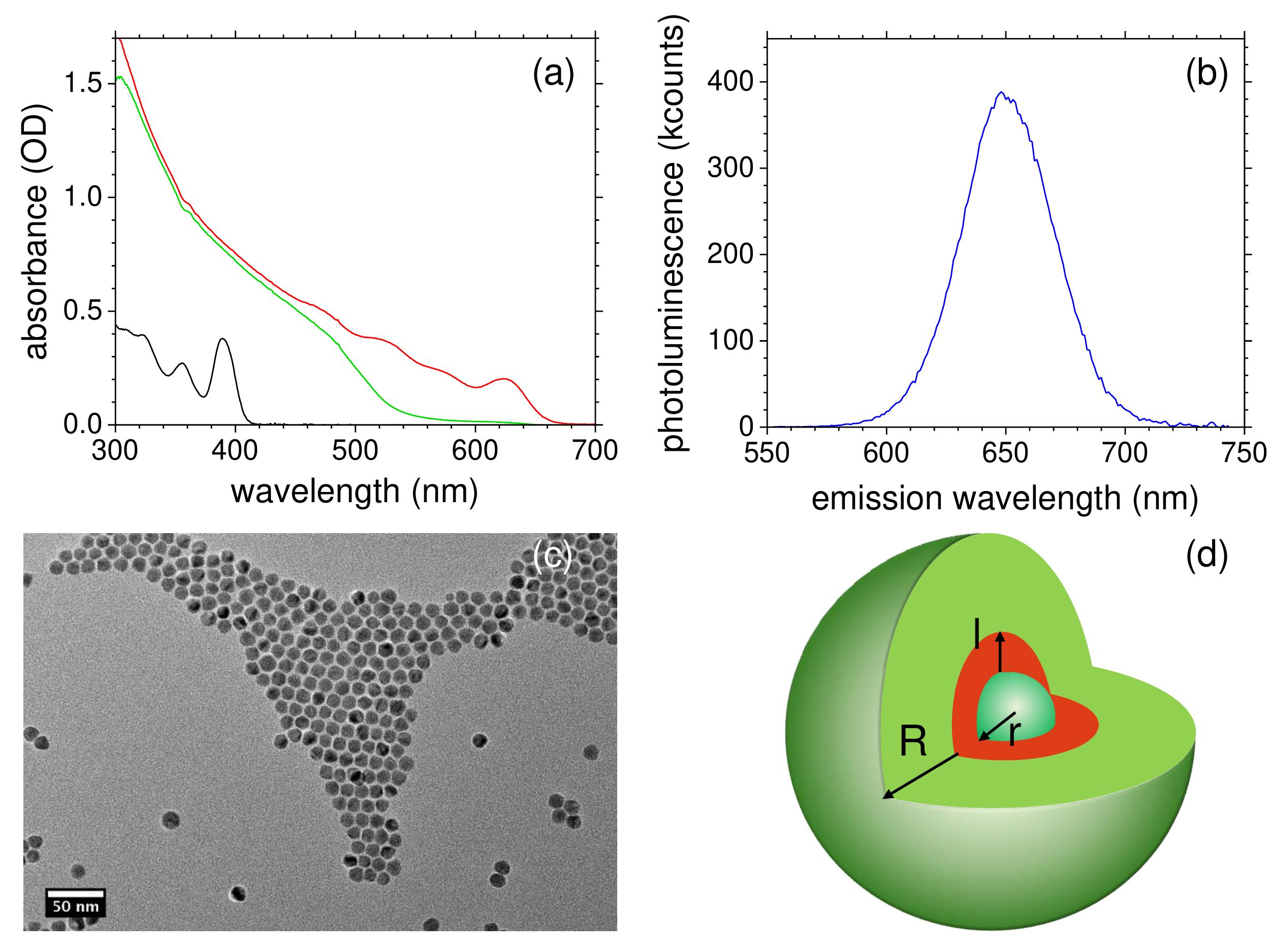}
\caption{(a) Absorption spectra of CdS core (black line), CdS/CdSe nanoshell (red line) and CdS/CdSe/CdS SQW (green line). (b) CdS/CdSe/CdS SQW emission spectrum. (c) TEM image of the synthesized spherical quantum wells. (d) Scheme of a single spherical quantum well structure:  a CdS (green) core of radius r=1.3\,nm, a CdSe (red) active layer of thickness l=1.7\,nm and a CdS (green) covering shell of thickness R=3.4\,nm.}
\label{fig.1}
\end{figure}
Details of the CdS/CdSe/CdS SQW synthesis are given in the Supplementary Material. Fig.~\ref{fig.1}\,(a) shows absorption spectra of an ensemble of CdS core (black), CdS/CdSe un-shelled SQWs (red) and CdS/CdSe/CdS (green) final SQWs. CdS cores present an absorption peak located at 390\,nm, pointing to a radius of $\rm r = 1.3$\,nm~\cite{jeong16acsna109297}. Absorption spectrum of CdS/CdSe SQWs show a first exciton located at 625\,nm, revealing a CdSe active layer thickness of $\rm l = 1.7$\,nm~\cite{jeong16acsna109297}. After the final CdS-shelling layer is added, the absorption spectra is dominated by the CdS absorption and a faint signature of the exciton is observed at 625\,nm. The PL signal shows a strong Gaussian-shaped resonance centered at 650\,nm, with a full width at half maximum (FWHM) of 46\,nm, fig.~\ref{fig.1}\,(b). We present in fig.~\ref{fig.1}\,(c) a transmission electron microscopy (TEM) image of the final SQWs. Single isolated particles are clearly visible on the TEM image. Statistics taken over 100 particles give an average SQW diameter of 12.7$\pm$1.2\,nm. This leads to the average particle schemed in fig.~\ref{fig.1}\,(d), with a CdS core of $\rm r = 1.3$\,nm, a CdSe active layer of $\rm l = 1.7$\,nm and a CdS shell thickness $\rm R = 3.4$\,nm.
\begin{figure}[t]
\centering
\includegraphics[width=8 cm]{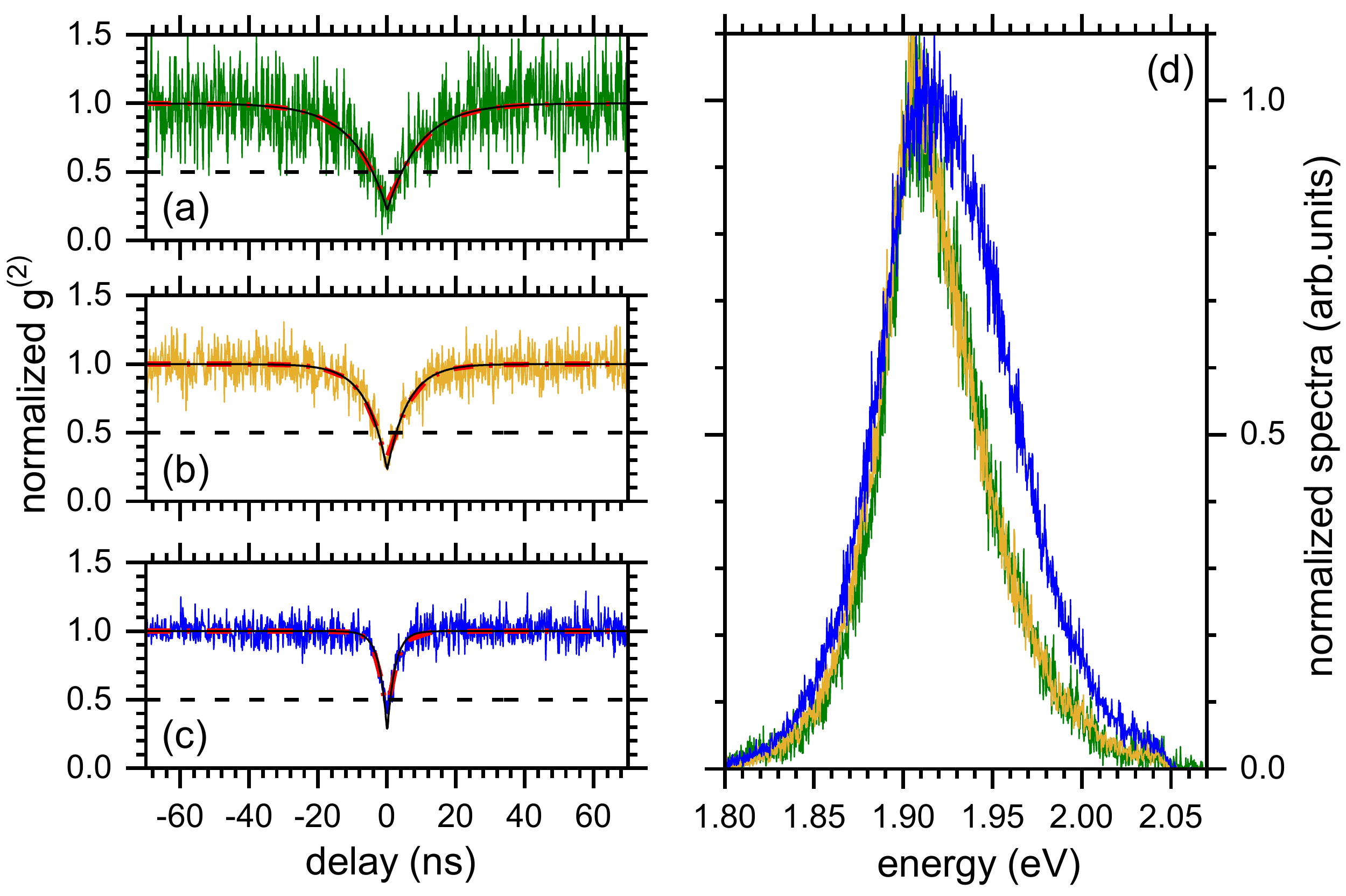}
\caption{(a), (b) and (c) Coincidence histogram at fast times recorded on a single SQW, at three different excitation intensities: $\rm I_l = 1$\,kW/cm$^2$ (green line), $\rm I_m = 3$\,kW/cm$^2$ (orange line) and $\rm I_h = 6.5$\,kW/cm$^2$ (blue line) respectively. Dashed-red lines in (a), (b) and (c) are  fitting results from a mono-exponential fit convoluted with the temporal response of the detectors. Black solid lines are the results of the un-convoluted (actual SQW antibunching) exponential component of the fit. The time constant obtained are $\rm \tau_{\rm l}$ = 9.3$\pm 0.6$\,ns, $\rm \tau_{\rm m}$ = 6.0$\pm$0.2\,ns and $\rm \tau_{\rm h}$ = 2.3$\pm$0.1\,ns for (a), (b) and (c) respectively. (d) Corresponding spectra recorded simultaneously with the antibunching curves of the single SQW probed in (a), (b) and (c), S$_{\rm l}$ (green), S$_{\rm m}$ (orange) and S$_{\rm h}$ (blue) respectively. Excitation wavelength was $\rm \lambda_{\rm exc}$ = 405\,nm.}
\label{fig.2}
\end{figure}
We report in fig.~\ref{fig.2}\,(a) coincidence measurements of the emission ($\rm g^{(2)}(t)$) of a single SQW (SQW1). Details of the experiments are presented in the Supplementary Material. The autocorrelation signal at zero time-delay in a coincidence setup $\rm g^{(2)}(t=0)$ appearing at less than 0.5 is the smoking-gun for identifying single-photon sources. Fig.~\ref{fig.2}\,(a) exhibits $\rm g^{(2)}(t=0) = 0.21$, indicating a high single-photon purity. This represents the first report of the antibunching of the light emitted by a single SQW, demonstrating their atom-like 3-dimensional confinement, contrary to what was argued so far~\cite{zamkov17JACS_1D_SQW,zamkov20nanoshell_bohr}. This $\rm g^{(2)}(t=0) = 0.21$ value is obtained from the black line in fig.~\ref{fig.2}\,(a), which represents an exponential fit of the data, using a home-made unbiased fitting algorithm (described elsewhere~\cite{baronnier21pccp}) taking into account the time-resolution of the setup (see Supplementary Material). The data were then fitted according to: 
\begin{equation}
\label{eq.1}
\rm g^{(2)}(t) = 1-Ce^{-\frac{|t|}{\tau_{\rm l,m,h}}}
\end{equation}
leading to $\rm \tau_{\rm l} = 9.3\pm$0.6\,ns, with $\rm C = 1-g^{(2)}(0)$ being the contrast at t = 0. The uncertainty on the fitting parameters was obtained by hundred bootstrap resampling of the experimental data and subsequent fitting~\cite{baronnier21pccp}. This measurement was performed at an excitation intensity of $\rm I_l = 1$\,kW/cm$^2$. We present in fig.~\ref{fig.2}\,(b) and (c) antibunching measurements at two more excitation intensities $\rm I_m = 3$\,kW/cm$^2$ (fig.~\ref{fig.2}\,(b)) and $\rm I_h = 6.5$\,kW/cm$^2$ (fig.~\ref{fig.2}\,(c)). The contrast at zero delay time in fig.~\ref{fig.2}(b) is $\rm g^{(2)}(0) = 0.23$, still well-below the 0.5 limit. The black line in fig.~\ref{fig.2}\,(b) represents a fit to the data following eq.~\ref{eq.1} and leads to a shorter time $\rm \tau_m = 6.0\pm0.2$\,ns. Probing the antibunching at an even higher intensity (fig.~\ref{fig.2}\,(c)) pushes the antibunching contrast at $\rm g^{(2)}(t=0) = 0.32$, still well-below the single photon limit of 0.5. We also note that the fit of the data gives a faster time $\rm \tau_h = 2.3\pm0.1$\,ns. These values obtained for the times which fit the best the data can be understood in the repumping rate framework~\cite{lounis00oct27}. The following expression for the repumping rate~\cite{lounis00oct27} was used: 
\begin{equation}
\label{eq.2}
\rm W_{\rm l,m,h} = \frac{I_{l,m,h}\sigma_{\rm SQW}\lambda_{\rm exc}}{hc} 
\end{equation}
Where $\sigma_{\rm SQW} = 5\times10^{-14}$\,cm$^2$ is the absorption cross section at an excitation wavelength $\rm \lambda_{exc} = $405\,nm of a single SQW (if we make the approximation that it is comparable to a sphere of CdS of diameter 13\,nm~\cite{klimov11PRL}), $\rm h$ is Planck constant and $\rm c$ is the speed of light in vacuum. Inserting these values in eq.~\ref{eq.2} This leads to $\rm W_l = 103$\,MHz, $\rm W_m = 309$\,MHz and $\rm W_h = 670$\,MHz, corresponding to repumping times $\rm \tau^r_l = 9.7$\,ns, $\rm \tau^r_m = 3.2$\,ns and $\rm \tau^r_h = 1.5$\,ns respectively. These values are comparable to those obtained from the fitting of data in fig.~\ref{eq.1}\,(a), (b) and (c) with eq.\,\ref{eq.1}. In the reprumping rate framework, the function which describes $\rm g^{(2)}(t)$ is~\cite{lounis00oct27}:
\begin{equation}
\label{eq.2b}
\rm g^{(2)}(t) = 1-Ce^{-(\frac{1}{\tau_{exc}} + W)|t|}
\end{equation}
Where $\rm \tau_{exc}$ and $\rm W$ are the decay time of the exciton in the SQW and $\rm W$ is the repumping rate. It is clear that if $\rm W\gg1/\tau_{exc}$, only the contribution due to the repumping rate will show-up in the $g^{(2)}(t)$ measurements. Our fitting results thus clearly indicate that the emission lifetime $\rm \tau_{exc}$ of SQW1 is slower than $\rm \tau^r_{l,m,h}$ and that we operate in the regime where the emission rate is dominated by the repumping rate. It is interesting to note that, even in the high intensity regime, where other reports on an ensemble of SQWs observe an emission dominated by the bi-exciton~\cite{nagamine20acsphot}, we clearly demonstrate that single photon emission remains in our case. We have recorded simultaneously to the $\rm g^{(2)}(t)$ measurements the corresponding emission spectra, presented in fig.~\ref{fig.2}\,(d). The spectra in green, orange and blue correspond to $\rm I_l$, $\rm I_m$ and $\rm I_h$ excitation intensities respectively. We see barely any difference between the low-intensity (green) and medium-intensity (orange) curves. On the other hand, we observe a clear difference in the shape of the emission between the high-intensity spectrum (blue) and the two others: a high-energy shoulder appears for E$\geq$1.9\,eV. Coupling this information, the decrease of the contrast at $\rm g^{(2)}(0)$ and previous works on ensemble of SQW~\cite{nagamine20acsphot,meng21nanoscale} it is tempting to assign that high-energy component to the contribution of the bi-exciton emission. To test this hypothesis, we have furthermore increased the excitation intensity on SQW1 up to $\rm I_{max} = 17$\,kW/cm$^2$ (see Supplementary Material): the emission remained unchanged, pointing to another physical process than the bi-exciton. We present in fig.~\ref{fig.3} intensity-dependent spectra performed on another single SQW (SQW2). Fig.~\ref{fig.3}\,(a) shows spectra recorded at increasing intensities from $\rm I_{l,2} = 1$\,kW/cm$^{2}$ (green), $\rm I_{m,2} = 4.4$\,kW/cm$^{2}$ (orange) and $\rm I_{h,2} = 17$\,kW/cm$^{2}$ (blue). As the intensity is increased, a high-energy shoulder rises as for SQW1, peaking at $\rm E_{2} = 1.925$\,eV. The energy shift is $\Delta E_2 =52$\,meV. Fig.~\ref{fig.3}\,(b) represents the curve at intensity $\rm I_{m,2} = 4.4$\,kW/cm$^{2}$ recorded before (orange) and after (black) having recorded the high-intensity spectrum at $\rm I_{h,2}$ (blue line in fig.~\ref{fig.3}\,(a)). This clearly shows that the process leading to the appearance of the high-energy component of the emission derives from an irreversible photo-induced process. We see that while both SQW1 and SQW2 have qualitatively similar behavior regarding high-intensity excitation, the quantitative shift of the high-energy transition can take different values. We have recorded this shift for 26 single SQWs and observed a moderate correlation (Pearson $\rm r = 0.36$) between the low-energy and high-energy emission energy (see Supplementary Material).  
\begin{figure}[t]
\centering
\includegraphics[width=8 cm]{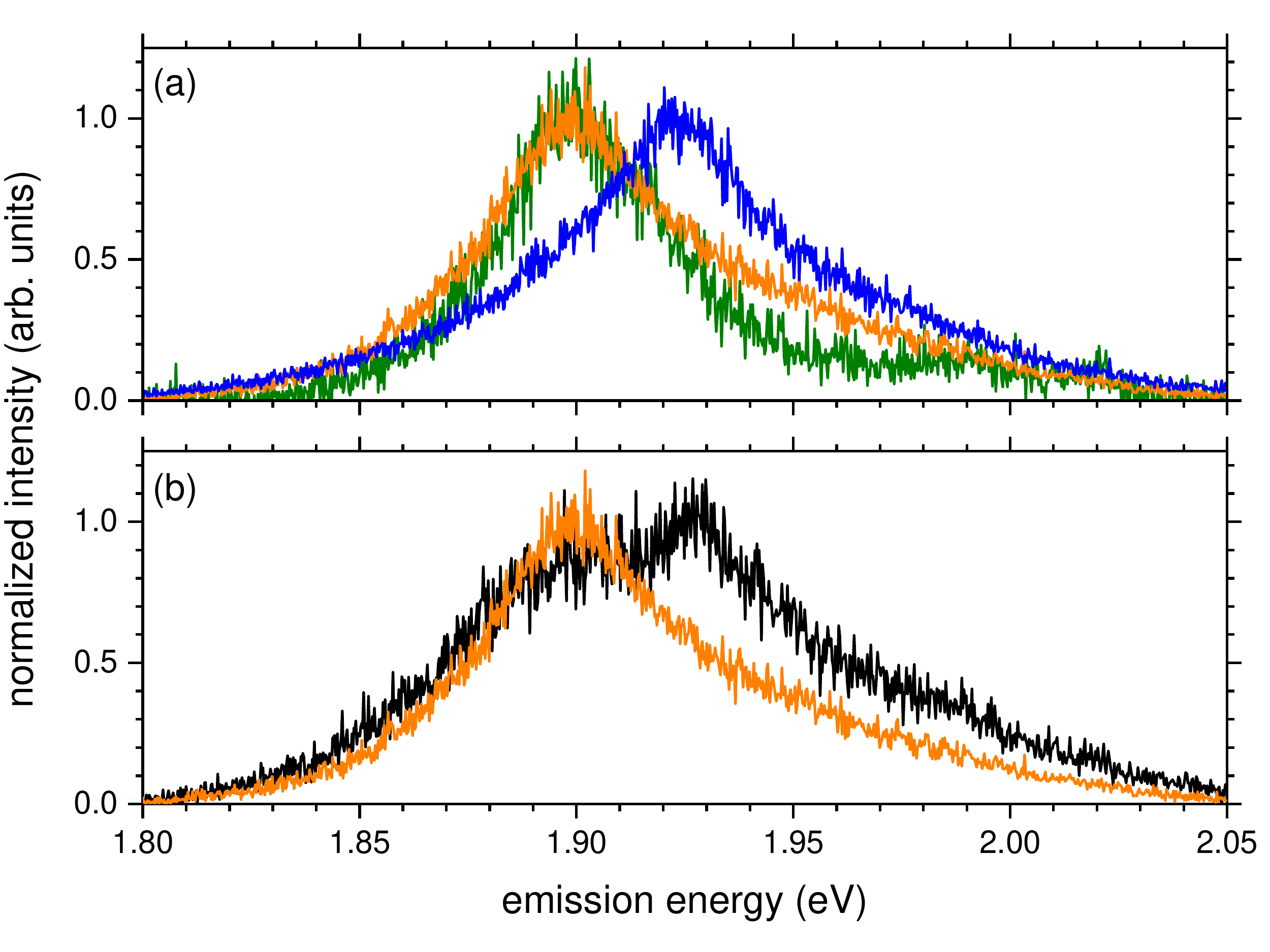}
\caption{(a): Spectra of SQW2 recorded at rising excitation intensities $\rm I_{l,2} = 1$\,kW/cm$^{2}$ (green), $\rm I_{m,2} = 4.4$\,kW/cm$^{2}$ (orange) and $\rm I_{h,2} = 17$\,kW/cm$^{2}$ (blue). (b): Spectra of SQW2 recorded at an excitation intensity $\rm I_{m,2} = 4.4$\,kW/cm$^{2}$ before (orange) and after (black) an excitation at $\rm I_{h,2} = 17$\,kW/cm$^{2}$. Excitation wavelength was $\lambda_{exc} = $405\,nm.}
\label{fig.3}
\end{figure}
\begin{table*}[t]
\caption{The main fit parameters of the decay models for 28 single SQWs. Listed are the four decay times $\tau_i$ and their respective contributions $f_i$ to the total signal (minus the constant background); uncertainties of the fit parameters were obtained by empirical standard deviation over the values obtained from the fits. The excitation intensity used for these measurements, was chosen to be large enough for reliable observation of all four decay components without saturation effects.}
\label{tab.1}
\begin{center}
\begin{tabular}{llccccccccccr}\hline
model & $\#$ SQWs & $\tau_1$ (ns) & f$_1$ ($\%$) & $\tau_2$\,(ns) & f$_2$ ($\%$)&  $\tau_3$\,(ns) & f$_3$ ($\%$) & $\tau_4$\,(ns) & f$_4$ ($\%$)\\ \hline
3-exp & 15 & 116$\pm$23 & 11$\pm$9& 40.7$\pm$4.7 & 80$\pm$12 &  10.2$\pm$4.8 & 9$\pm$6 & n.a. & n.a.\\
4-exp & 13 & 60.3$\pm$10.3 & 44$\pm$19 & 31.5$\pm$6.5 & 44$\pm$20 & 7.4$\pm$1.8 & 7.2$\pm$6.3 & 466$\pm$370 & 4.8$\pm$2.8 \\ \hline
\end{tabular}
\end{center}
\end{table*}
\begin{figure}[t]
\centering
\includegraphics[width=8 cm]{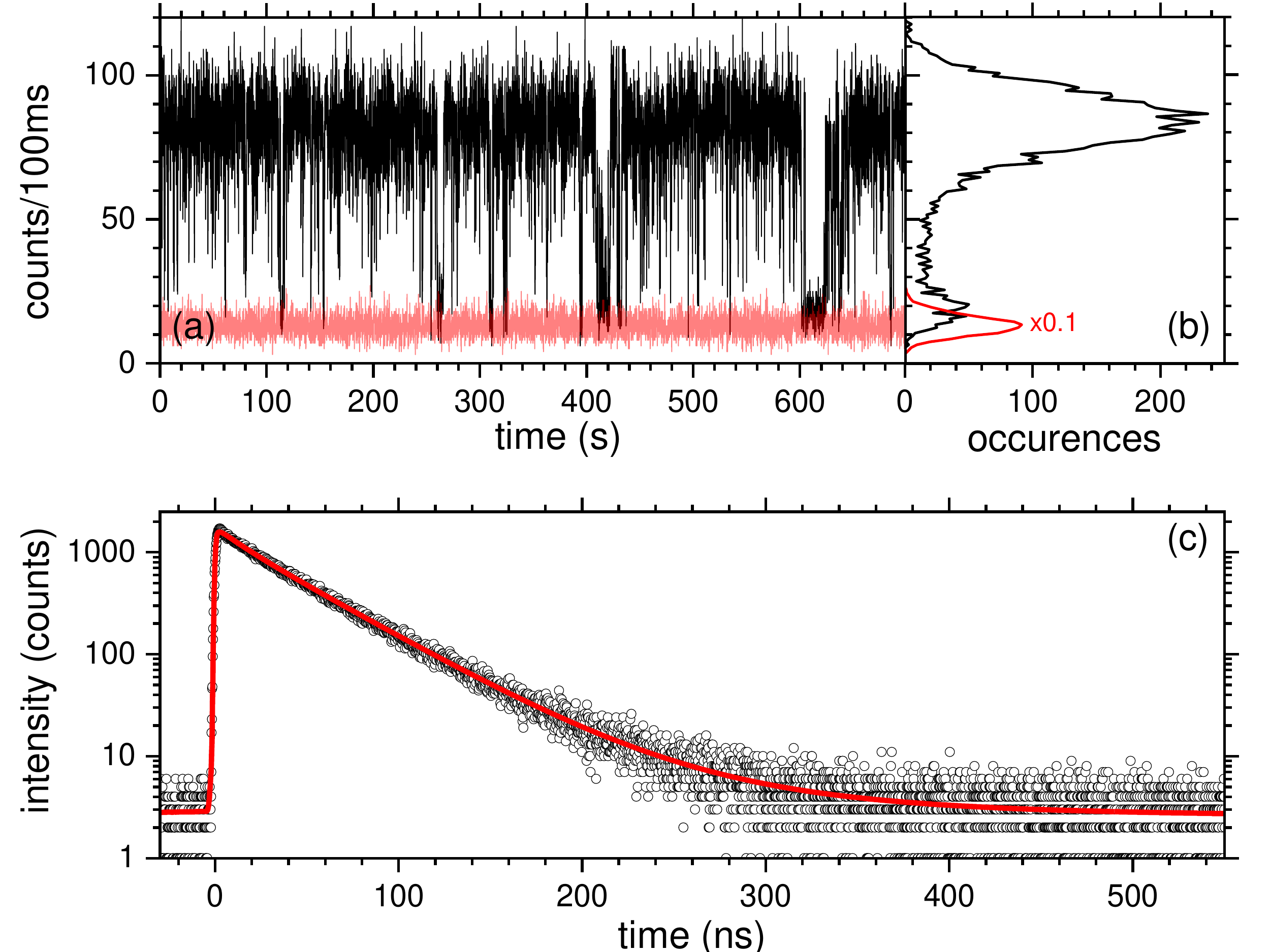}
\caption{(a): Intensity timetrace measured on a single spherical quantum well (black). Dark noise of the apparatus (red) (b) Intensity count distribution exhibiting a two-level distribution typical of blinking behavior (black). A bi-Gaussian fit (not shown) permits us to conclude that the SQW spends 85\,$\%$ of its emission time in the bright state. Red line represents the ounts distribution of the dark noise. (c) Emission lifetime recorded on the SQW from (a) (dark-white circles) and a 3-exponential fit (red line) to describe its behavior. The best-fit lifetimes are found to be $\rm \tau_1 = 12.1\pm2.3$\,ns, $\rm \tau_2 = 41.9\pm1.0$\,ns and $\rm \tau_3 = 107.7\pm26.1$\,ns. Data were recorded at an excitation intensity of $\rm I = 200$\,W/cm$^2$, a repetition rate of 2\,MHz and an excitation wavelength of $\lambda_{\rm exc}$ = 450\,nm.}
\label{fig.4}
\end{figure}
We present in fig.~\ref{fig.4}\,(a) an intensity timetrace recorded on a single SQW (SQW3) with an excitation intensity $\rm I = 200$\,W/cm$^2$. The timetrace exhibits a fast blinking~\cite{nirmal96oct31}, i.e. a succession of sequences of emissive (bright) and non-emissive (dark) states, with intra-time-bin events, and the longest dark event lasting less than 20 seconds over the 700 seconds experimental time. We plot in fig.~\ref{fig.4}\,(b) the projection of the counts from fig.~\ref{fig.4}\,(a). It shows a bimodal distribution which can be fitted with two Gaussians (not shown): one for the bright state, one for the dim state. The following ratio gives the fraction of the time the SQW spends in the bright state:
\begin{equation}
\label{eq.5}
\rm \alpha_{\rm bright} = \frac{\rm I_{\rm bright}}{\rm I_{\rm bright} + I_{\rm dim}}
\end{equation}
where $\rm I_{bright}$ and $\rm I_{dim}$ are the area of the two Gaussian used to fit the bright and dim count-distributions respectively. We find $\rm \alpha_{bright} = 85$\,$\%$ for SQW3. We have repeated this study over 28 single SQWs. We find that they spend a median 89\,$\%$ of their emissive time in the bright state, as previously observed with less statistics~\cite{jeong16acsna109297}. Fig.~\ref{fig.4}\,(c) presents the decay curve obtained on SQW3 (black circle) at the excitation intensity $\rm I = 200$\,W/cm$^2$, together with the corresponding fit (red line) obtained with the same home-made fitting algorithm used to fit the $\rm g^{(2)}(t)$ curves in fig.~\ref{fig.2}~\cite{baronnier21pccp}. The fitting method allows us to determine the number of exponential functions that allow for the description of the experimental data with the help of a statistical test returning a p-value~\cite{baronnier21pccp}. The model used for the N-component intensity as a function of time was as follow:
\begin{equation}
\rm I_N(t) = y_0 + \Sigma^N_{i=1} A_i e^{-t/\tau_i}  
\label{eq7}
\end{equation}
Where $\rm y_0$, $\rm A_i$ and $\rm \tau_i$ are the background signal, the amplitude and decay time constant of the $\rm i^{th}$ component respectively. We find for SQW3, that three exponential functions (plus a background) are necessary to describe appropriately the data, but adding a forth component does not ameliorate the results with a high-enough statistical significance: $\rm p = 10^{-15}$ between N=2 and N=3 exponential, and $\rm p = 0.2$ between N=3 and N=4 exponential (any p-value $\rm p>$0.05 rejects the null hypothesis). This indicates that adding a fourth component to the fit does not describes any meaningful physical process. The lifetimes found as the best-fit parameters are $\rm \tau_1 = 107.7\pm26.1$\,ns, $\rm \tau_2 = 41.9\pm1.0$\,ns and $\rm \tau_3 = 12.1\pm2.3$\,ns. The first component $\rm \tau_1$ encompasses 4.7$\pm$2.3\,$\%$ of the total emitted photons, while 90.4$\pm$1.7\,$\%$ are attached to the component $\rm \tau_2$ and 4.9$\pm$1.6\,$\%$ corresponds to $\rm \tau_3$. The uncertainties are obtained $via$ hundred bootstrap resampling of the data and subsequent fitting. It is thus clearly seen that the large majority of the photons are emitted in the second component $\rm \tau_2$. This is coherent with the findings in fig.~\ref{fig.4}\,(b) which suggest the existence of a dominant bright state. The lifetime-statistics obtained on the 28 single SQWs is presented in Table~\ref{tab.1} and demonstrates two different class of behavior: 15 SQWs obey a N=3 exponential dynamics, while 13 are better described by a N=4 exponential behavior. Those presenting a N=3 exponential behavior have a dynamic comparable to SQW3 presented in fig.~\ref{fig.4}, with a clear dominance of the second emissive state which shows an average value $\rm \tau_2 = 40.7\pm 4.7$\,ns, containing on average $80\pm 12\%$ of the photons. On the other hand, the SQWs obeying a N=4 exponential emission dynamics see the addition of a slow ($\rm \tau_4 = 466\pm370$\,ns) fourth component containing on average $\rm 4.8\pm 2.8\,\%$ of the detected photons. The first two components average at $\rm \tau_1 = 60.3\pm10.3$\,ns and $\rm \tau_2 = 31.5\pm6.5$\,ns, accumulating 44$\pm$19\,$\%$ and 44$\pm$20\,$\%$ respectively, thus accounting for almost 90\,$\%$ of the emitted photons. The experiments in~\cite{jeong16acsna109297} reported single exponential decays for ensemble measurements, attributed to the neutral exciton X$_{\rm 0}$. We already found that a much richer dynamic is present in an ensemble of SQWs~\cite{meng21nanoscale}. However, the antibunching highlighted in fig.~\ref{fig.2}, which does not vanishes with increasing power, coupled with the absence of a fast ($<$1\,ns) recombination lifetime, lead us to conclude that none of the above transitions come from the bi-exciton. It thus means that only linear excitons contribute to the decay. We attribute the time $\rm \tau_1$ in both N=3 and N=4 exponential model to X$_{\rm 0}$. The time $\rm \tau_2$ is attributed to the negatively charged exciton (X$^-$) and $\rm \tau_3$ to the positively charged exciton (X$^+$). The time $\rm \tau_4$ being too long to be attributed to a direct exciton transition, it must be attributed to recombination $via$ surface defects. In this allocation of the transitions, the quantum yield (QY) of X$_0$ and X$^-$ are similar for the SQWs obeying the N=4 exponential dynamics: $\rm \tau_1 \sim 2\tau_2$, as expected because X$^-$ has twice more possible radiative decay paths compared to X$_0$~\cite{guyotsionnest08crphy9777}. On the other hand, the QY of X$^-$ is lowered compared to that of X$_0$ for the SQWs following the 3-exponential dynamics, since $\rm \tau_1 \sim 3\tau_2$. Another difference is that most of the photons (80\,$\%$) are emitted by te X$^-$ for N=3 exponential SQWs, while the photons are evenly distributed through X$_0$ (44\,$\%$) and X$^-$ (44\,$\%$) for the N=4 exponential SQWs. The positively charged exciton X$^+$ lifetime $\rm \tau_3$ should actually lead to the same lifetime as X$^-$ ($\rm \tau_2$) since the additional radiative decay paths are the same as for $\rm X^-$. However, it is well-documented that in CdSe-base emissive NCs X$^+$ has an altered, but non-zero, quantum yield compared to X$^-$~\cite{mahler08natma7659,guyotsionnest08crphy9777}, leading to a faster decay time. Those interpretations are also coherent with the results obtained in fig.~\ref{fig.4}\,(b). The maximum of the emission is centered at 84\, counts/100\,ms. A single emitting level should lead to a theoretical standard deviation of the Poissonian distribution of $\sigma_{\rm th} = 9$\,counts/100\,ms. However, we find an experimental value of $\sigma_{\rm exp} = 1.2\times\sigma_{\rm th}$. This is either the signature of the sub-timebin blinking or several emissive entities with slightly different quantum yield, $\rm X_0$ and $\rm X^-$. The dimmed-state has a lower emission QY, but non-zero considering the offset between its count distribution average (19 counts/100\,ms) and the background counts of the system (13 counts/100\,ms). That gives a non-zero, emission of the dimmed state of 6 counts/100\,ms. In the framework of that assignment of the emissive components, we can deduce the QY of X$^+$ emission relative to that of X$^-$, in both class of SQWs: those with N=3 or N=4 exponential dynamics. We define the QY of X$^+$ as~\cite{aubret16nanos82317}:
\begin{equation}
\centering
\eta_{X^+} = \frac{k_{3,r}}{k_{3,r}+k_{3,nr}}
\label{eq.3}    
\end{equation}
where $\rm k_{3,r}$ and $\rm k_{3,nr}$ are the radiative and non-radiative decay rates of $\rm X^+$ respectively, and using: 
\begin{equation}
\centering
\frac{1}{\tau_3} = \frac{1}{\tau_2} + \frac{1}{\tau_{3,nr}}
\label{eq.4}    
\end{equation}
where $\rm k_{3,r} = 1/\tau_2$ and $\rm k_{3,nr} = \frac{1}{\tau_{3,nr}}$. Using eq.~\ref{eq.3} and~\ref{eq.4}, and the average values given in Table~\ref{tab.1}, we find a similar emission quantum yield $\eta^{3exp}_{X^+}$ = 25\,$\%$ and $\eta^{4exp}_{X^+}$ = 24\,$\%$ for both class of SQWs.
\section{Conclusion}
In conclusion, we have synthesized CdS/CdSe/CdS spherical quantum wells emitting at 650\,nm and shown that they are single photon emitters by reporting the first antibunching of their emission. This has proven their confinement to be 3-dimensional. We have further demonstrated that single photon emission remains, even at high intensities where the antibunching is dominated by the repumping rate. intensity-dependent spectral measurements, recorded simultaneously with the antibunching, exhibit an irreversible blue-shifted emission component that is not attributed to the bi-exciton, but to a photo-aging process. Statistical characterization of that blue-shift over 26 single SQWs have shown a moderate correlation between the low-intensity emission and the blue-shifted component with a Peasron correlation coefficient $\rm r = 0.36$. Intensity timetraces recorded over 28 single SQWs have been conducted and it has shown a low 2-level blinking behavior. Their median time spent in the bright emitting state was found to be 89\,$\%$. We have performed lifetime measurements on those 28 single SQWs, coupled to a home-made fitting routine that includes a statistical test, and demonstrated that the SQWs obey either to a 3- or 4-component dynamic. In both cases, we speculated that emission is structured by that of $\rm X_0$, $\rm X^-$ and $\rm X^+$ (3-component) plus the long lived emission from defects (4-component). We found that emission is mostly (80\,$\%$) from $X^-$ in the 3-component class of SQWs, while it is evenly distributed over $\rm X_0$ (44\,$\%$) and $\rm X^-$ (44\,$\%$) for the SQWs obeying the 4-component dynamic. From those results, we have identified the dim state in the intensity timetraces to come from $\rm X^+$, and we deduced its emission quantum yield relative to that of $\rm X^-$ to be $\sim$25\,$\%$ for both class of SQWs. Those results demonstrate that SQWs can be used as the single photon sources and give the most extensive, so far, characterization of their optical properties at the single particle level. 
\acknowledgments
The authors thank J.-F. Sivignon, Y. Guillin, G. Montagne and the Lyon center for nanoopto
technologies (NanOpTec) for technical support. This work was financially supported
by the Agence Nationale de Recherche (ANR-16-CE24-0002), IDEXLYON of Universit{\'e} de
Lyon in the framework “Investissement d’Avenir” (ANR-IDEX-0005) and the F{\'e}d{\'e}ration de Recherche Andr´e–Marie Amp{\`e}re (FRAMA).
\section{Supporting Information Available}
Supporting information on the synthesis of the spherical quantum wells and high intensity spectra is available online free of charge.

\end{document}